\documentclass[aps,pre,twocolumn,groupedaddress,floatfix]{revtex4-1}
\pdfoutput=1

\usepackage{natbib}
\usepackage{graphicx}
\usepackage{amssymb,amsfonts,amsmath}

\newcommand{\eq}[1]{Eq.~(\ref{#1})}
\newcommand{\Eq}[1]{Equation~(\ref{#1})}
\newcommand{\fig}[1]{Fig.~\ref{#1}}
\newcommand{\Fig}[1]{Figure~\ref{#1}}
\newcommand{\eqs}[2]{Eqs.~(\ref{#1}) and (\ref{#2})}
\newcommand{\figs}[2]{Figs.~\ref{#1} and \ref{#2}}

\newcommand{\citeref}[1]{Ref.~\citep{#1}}

\def \Ns{Ns}

\def \T2{T_2}

\begin{document}

\title{Interference limits resolution of selection pressures from linked neutral diversity}
\author{Benjamin H. Good$^{1}$}
\author{Aleksandra M. Walczak$^{2}$}
\author{Richard A. Neher$^{3}$}
\author{Michael M. Desai$^{1}$}
\affiliation{\mbox{${}^1$ Departments of Organismic and Evolutionary Biology and of Physics and FAS Center for Systems Biology, Harvard University}}
\affiliation{\mbox{${}^2$ CNRS-Laboratoire de Physique Th\'eorique de l'\'Ecole Normale Sup\'erieure}}
\affiliation{\mbox{${}^3$ Max Planck Institute for Developmental Biology, T\"ubingen, Germany}}

\begin{abstract}
Pervasive natural selection can strongly influence observed patterns of genetic variation, but these effects remain poorly understood when multiple selected variants segregate in nearby regions of the genome. Classical population genetics fails to account for interference between linked mutations, which grows increasingly severe as the density of selected polymorphisms increases. Here, we describe a simple limit that emerges when interference is common, in which the fitness effects of individual mutations play a relatively minor role. Instead, molecular evolution is determined by the variance in fitness within the population, defined over an effectively asexual segment of the genome (a ``linkage block''). We exploit this insensitivity in a new ``coarse-grained'' coalescent framework, which approximates the effects of many weakly selected mutations with a smaller number of strongly selected mutations with the same variance in fitness. This approximation generates accurate and efficient predictions for the genetic diversity that cannot be summarized by a simple reduction in effective population size. However, these results suggest a fundamental limit on our ability to resolve individual selection pressures from contemporary sequence data alone, since a wide range of parameters yield nearly identical patterns of sequence variability.
\end{abstract}

\date{\today}
\maketitle

Natural selection maintains existing function and drives adaptation, altering patterns of diversity at the genetic level. Evidence from microbial evolution experiments \citep{barrick:etal:2009} and natural populations of nematodes \citep{andersen:etal:2012}, fruit flies \citep{sella:etal:2009}, and humans \citep{lohmueller:etal:2011} suggests that selection is common and that it can impact diversity on genome-wide scales. Understanding these patterns is crucial, not only for studying selection itself, but also for inference of confounded factors such as demography or population structure. However, existing theory struggles to predict genetic diversity when many sites experience selection at the same time, which limits our ability to interpret variation in DNA sequence data.

Selection on individual nucleotides can be modeled very precisely, provided that the sites evolve in isolation. But as soon as they are linked together on a chromosome, selection creates correlations between nucleotides that are difficult to disentangle from each other. This gives rise to a complicated many-body problem, where even putatively neutral sites feel the effects of selection on nearby regions. Many authors neglect these correlations, or assume that they are equivalent to a reduction in the effective population size, so that individual sites evolve independently \citep{charlesworth:2009}. This assumption underlies several popular methods for inferring selective pressures and demographic history directly from genetic diversity data \citep{sawyer:hartl:1992, williamson:etal:2005, keightley:eyre-walker:2007, boyko:etal:2008}. Yet there is also extensive literature [recently reviewed in \citep{neher:2013}] which shows how the independent sites assumption breaks down when the chromosome is densely populated with selected sites. When this occurs, the fitness effects and demographic changes inferred by these earlier methods become increasingly inaccurate \citep{bustamante:etal:2001, messer:petrov:2013}.

Linkage plays a more prominent role in models of \emph{background selection} \citep{charlesworth:etal:1993} and \emph{genetic hitchhiking} \citep{maynard-smith:haigh:1974}, which arise in the limit of strong negative and strong positive selection respectively. Although initially formulated for a two-site chromosome, both can be extended to larger genomes as long as the selected sites are sufficiently rare that they can still be treated independently. Simple analytical formulae can be derived in this limit, motivating extensive efforts to distinguish signatures of background selection and hitchhiking from sequence variability in natural populations \citep{cutter:payseur:2013}. However, this data has uncovered many instances where selection is neither as rare nor as strong as these simple models require \citep{barraclough:etal:2007, bartolome:charlesworth:2006, betancourt:etal:2009, lohmueller:etal:2011, seger:etal:2010, subramanian:2012, ofallon:2013}. Instead, substantial numbers of selected polymorphisms segregate in the population at the same time, and these mutations interfere with each other as they travel towards fixation or loss. The genetic diversity in this \emph{interference selection} \citep{comeron:kreitman:2002} or \emph{weak Hill-Robertson interference} \citep{mcvean:charlesworth:2000} regime is poorly understood, especially in comparison to background selection or genetic hitchhiking. The qualitative behavior has been extensively studied in simulation \citep{przeworski:etal:1999, mcvean:charlesworth:2000, comeron:kreitman:2002, williamson:orive:2002, seger:etal:2010, messer:petrov:2013}, and this has led to a complex picture in which genetic drift and chance associations between mutations cause rapid changes in allele frequency. In principle, these simulations can also be used for inference or model comparison using approximate likelihood methods \citep{lohmueller:etal:2011}, but in practice, performance concerns severely limit both the size of the parameter space and the properties of the data that can be analyzed in this way.

Here, we will show that in spite of the complexity observed in earlier studies, simple behaviors do emerge when interference is sufficiently common. When fitness differences are composed of many individual mutations, we obtain a type of central limit theorem, in which diversity at putatively neutral sites is determined primarily by the variance in fitness within the population over a local, \emph{effectively asexual} segment of the genome. We exploit this simplification to establish a coalescent framework for generating predictions under interference selection, which is based on a \emph{coarse-grained}, effective selection strength and effective mutation rate. This leads to accurate and efficient predictions for a regime that is often implicated in empirical data, but has so far been difficult to establish more rigorously. In addition, these findings suggest a fundamental limit on our ability to infer selection pressures from patterns of linked neutral variability, even in an ideal setting. When interference is widespread, neutral diversity can no longer be used to distinguish between a small number of strongly selected mutations and a much larger number of weakly selected mutations, which has important implications for the interpretation of sequence data in this regime.

\section{Model}

\begin{figure}[t]
\centering
\includegraphics[width=1.0\columnwidth]{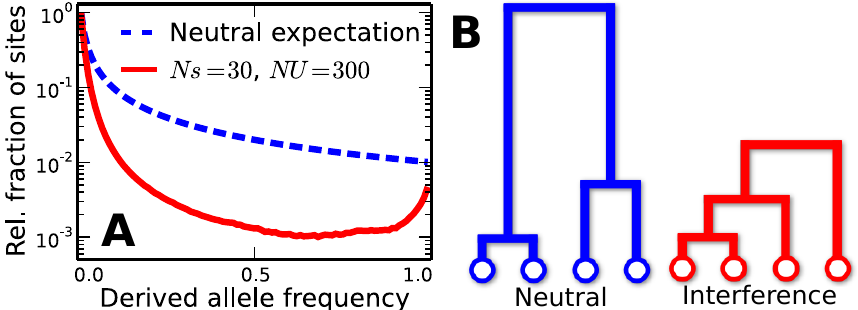}
\caption{(A) A typical silent site frequency spectrum under pervasive interference selection. The solid red line is averaged over many independent simulations of our simple purifying selection scenario with $Ns=30$, $NU=300$, $R \approx 0$, and a sample size of $n=100$ chromosomes. For comparison, the neutral expectation is given by the dashed blue line. (B) A schematic illustration of the genealogical structure observed in neutral populations (left) and those subject to widespread interference (right). These genealogical distortions give rise to the differences in the site frequency spectrum. \label{fig:frequency-spectrum-diagram}}
\end{figure}

We investigate the effects of widespread selection in the context of a simple and well-studied model of molecular evolution. Specifically, we consider a population of $N$ haploid individuals, each of which contains a single linear chromosome that accumulates mutations at a total rate $U$ and undergoes crossover recombination at a total rate $R$. We assume that the genome is sufficiently large, and epistasis is sufficiently weak, that the fitness contribution from each mutation is drawn from some distribution of fitness effects $\rho(s)$ which remains constant over the relevant time interval. For the sake of concreteness and connection with previous literature, we will focus on the special case where all mutations confer the same deleterious fitness effect $-s$, which approximates a potentially common scenario where a well-adapted population is subject to purifying selection at a large number of sites. However, our results will hold for more general distributions of fitness effects, both beneficial and deleterious, provided that individual mutations are sufficiently weak or the overall mutation rate is sufficiently large. Finally, since the effects of linked selection are most pronounced in regions of low recombination, we devote the bulk of our analysis to the asexual limit where $R \approx 0$. Later, we will show that recombining genomes can be treated as an extension of this limit by means of an appropriately defined \emph{linked  block}, within which recombination can be neglected.

These assumptions define a ``null-model'' of sequence evolution with a straightforward computational implementation (SI Appendix). Yet a quantitative understanding of this model remains elusive for many biologically relevant parameters. Even in the simple case where mutations share the same deleterious fitness effect, analytical predictions are limited to combinations of $N$, $U$, $s$, and $R$ such that the density of selected polymorphisms is small. As $Ns \to \infty$, these populations converge to the well-known \emph{background selection limit} \citep{hudson:kaplan:1995}, where molecular evolution is identical to that of a neutral population with an effective population size,
\begin{align}
\label{eq:bgs-limit}
N_e = N e^{- 2U/(2s+R) } \, .
\end{align}
For finite $Ns$, non-neutral corrections can be calculated using structured coalescent methods as long as $N_e s$ is sufficiently large that \eq{eq:bgs-limit} provides the dominant contribution \citep{hudson:kaplan:1994, gordo:etal:2002, zeng:charlesworth:2011, walczak:etal:2012, nicolaisen:desai:2012}. In the present work, we focus on the opposite extreme, the so-called \emph{interference selection regime}, where mutation rates are sufficiently high or fitness effects sufficiently weak that many selected polymorphisms segregate in the population at once. In this regime, structured coalescent methods break down \citep{gordo:etal:2002, walczak:etal:2012}.

\begin{figure*}[t]
\centering
\includegraphics[width=0.95\textwidth]{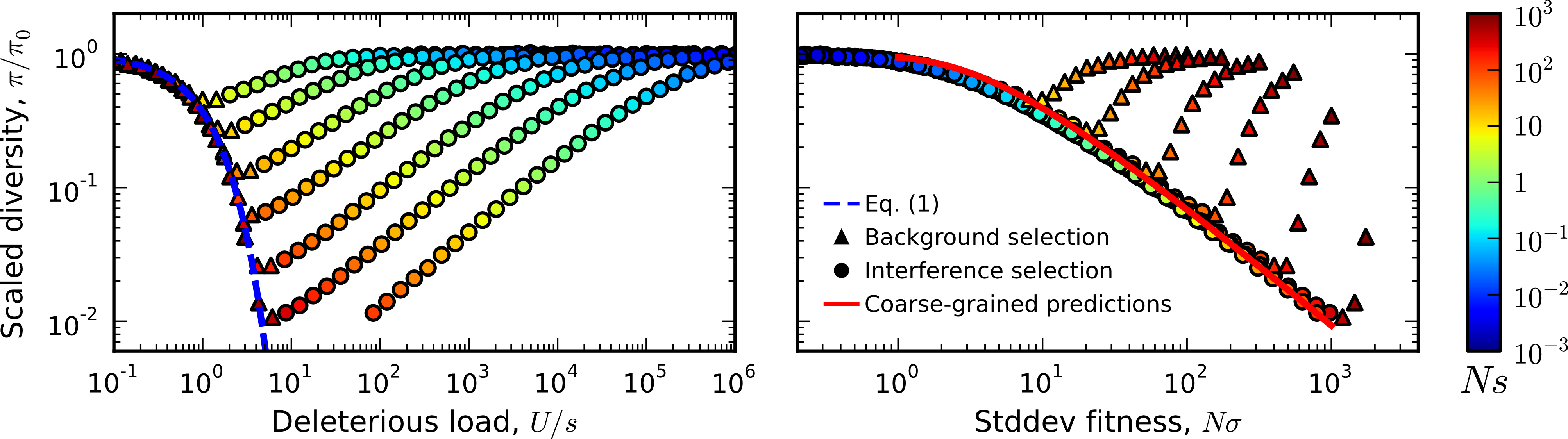}
\caption{The average reduction in silent site diversity relative to the neutral expectation $\pi_0 \propto N$. Colored points are measured from forward-time simulations of the simple purifying selection scenario in \fig{fig:frequency-spectrum-diagram} for $Ns \in (10^{-3},10^{3})$ and $NU=10,30,100,300,1000,3000,10000$. Triangles and circles distinguish populations that are classified into the ``background selection'' and ``interference selection'' regimes, respectively, according to the criteria outlined in the Analysis. In the left panel, these results are plotted as a function of the deleterious load $\lambda = U/s$, and the background selection prediction from \eq{eq:bgs-limit} is given by the dashed line. The right panel shows the same set of results plotted as a function of the observed standard deviation in fitness, and the solid line denotes the ``coarse-grained'' predictions described in the Analysis. Note that for populations in the background selection regime (triangles), $\pi/\pi_0$ is determined primarily by the deleterious load, independent of $Ns$ and $NU$. For populations in the interference selection regime (circles), $\pi/\pi_0$ is determined primarily by the standard deviation in fitness. \label{fig:t2-collapse} }
\end{figure*}

\begin{figure}[h!]
\centering
\includegraphics[width=0.92\columnwidth]{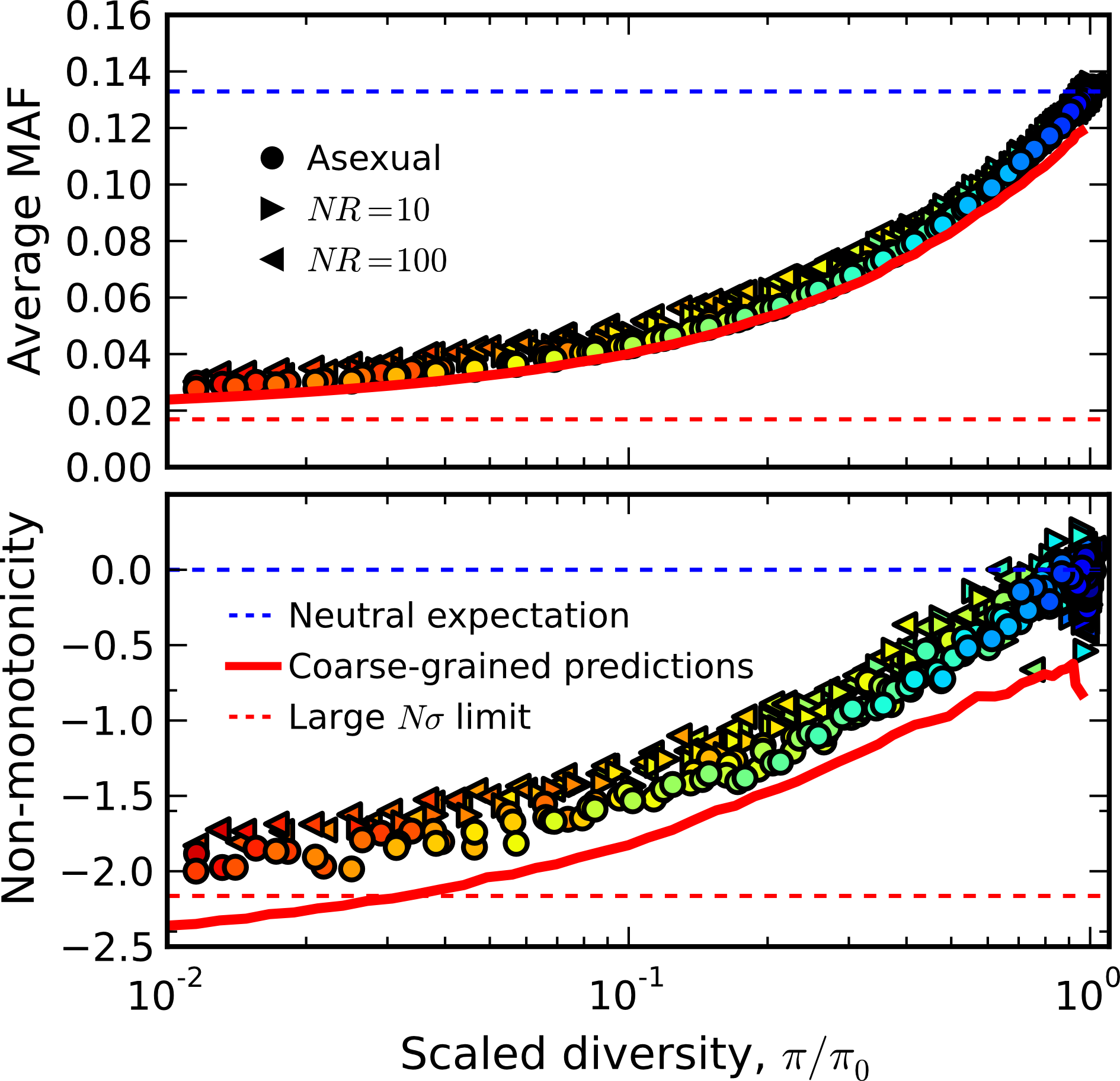}
\vspace{0.5em}
\caption{Signatures of pervasive interference selection in the silent site frequency spectrum for a sample of $n=100$ individuals. (A) An excess of rare alleles measured by the average minor allele frequency. (B) Non-monotonic or ``U-shaped'' behavior at high frequencies measured by $\log \left[ \min_i S_n(i) / S_n(n-1) \right]$. Both statistics are plotted as a function of the reduction in pairwise diversity,  $\pi/\pi_{0}$. Circles denote the subset of simulations in \fig{fig:t2-collapse} that were classified into the interference selection regime, while the right- and left-pointing triangles depict an analogous set of simulations for recombining genomes with $NR = 10$ and $NR=100$, respectively. All points are colored according to the same scale as \fig{fig:t2-collapse}. For comparison, the solid red lines show the ``coarse-grained'' predictions described in the Analysis, while the dashed lines show the corresponding predictions under neutrality (blue) and for the large $N\sigma$ limit in \citeref{neher:hallatschek:2013} (red). \label{fig:nonparametric-stats-collapse}}
\end{figure}

For concreteness, we concentrate on the genetic diversity at an unconstrained locus (e.g., a silent or synonymous site) embedded near the center of the chromosome. We focus in particular on the information contained in the polymorphic site frequency spectrum, $S_n(i)$, which counts the number of polymorphic sites shared by $i=1,\ldots,n-1$ individuals in a sample of size $n$. A typical example in the presence of interference is illustrated in \fig{fig:frequency-spectrum-diagram}A, which uses parameters consistent with the fourth (dot) chromosome of \emph{D. melanogaster} (SI Appendix). Apart from a reduction in the overall level of polymorphism, the most prominent features of this frequency spectrum include an excess of rare alleles similar to that observed after a recent population expansion \citep{kaiser:charlesworth:2008, seger:etal:2010}, and a non-monotonic (or ``U-shaped'') dependence at high frequencies\footnote{Due to the simultaneous enrichment of rare alleles, this nonmonotonicity does \emph{not} imply an excess of common variants over the neutral expectation. Indeed, Fay and Wu's $H$ \citep{fay:wu:2000} remains positive.}  \citep{neher:hallatschek:2013}, which is more commonly associated with selective sweeps \citep{fay:wu:2000} or complex demographic structure \citep{wakeley:aliacar:2001}. Since the mutations at this locus are neutral, these distortions in the site frequency spectrum ultimately reflect distortions in the genealogy of the sample from the predictions of neutral coalescent theory (\fig{fig:frequency-spectrum-diagram}B). The excess of rare alleles is due to an increase in the relative length of recent branches, compared to more ancient ones, and the non-monotonic behavior manifests itself in measures of genealogical imbalance \citep{seger:etal:2010}. Together, these features provide strong evidence for selection that could be exploited for inference in natural populations. But apart from the computationally-costly simulations above, there are few available methods for predicting genealogies or site frequency spectra when many selected mutations segregate in the population at once.

In the following section, we use simulations of our evolutionary model to show that widespread interference generates striking regularities in the patterns of genetic diversity. This approach is similar to earlier simulation studies, but with the express purpose of identifying patterns that can be exploited for \emph{prediction}, rather than simply describing the behavior observed in the presence of interference. In the Analysis, we generalize these patterns and use them to establish a new coalescent framework for predicting genetic diversity when interference is common. 

\section{Results}

Using the forward-time simulations described in SI Appendix, we measured the average site frequency spectrum for 280 asexual populations in our simple purifying selection model, where all mutations share the same deleterious fitness effect. Mutation rates ($NU$) ranged from $10$ to $10^4$ and selection strengths ($Ns$) ranged from $10^{-3}$ to $10^3$. Depending on the underlying parameters, we partitioned these populations into ``background selection'' or ``interference selection'' regimes using the criteria outlined in the Analysis. Loosely speaking, the background selection regime corresponds to the region where the structured coalescent is valid. \Fig{fig:t2-collapse} shows the reduction in diversity observed in our simulations, as measured by the pairwise heterozygosity $\pi$ relative to its neutral expectation, $\pi_0 \propto N$. As expected, the reduction in diversity is well-approximated by \eq{eq:bgs-limit} in the background selection regime (triangle symbols) \citep{gordo:etal:2002}, but it breaks down for populations in the interference selection regime (circles) \citep{kaiser:charlesworth:2008}. In addition, the traditional measure of the deleterious load $\lambda = U/s$ ceases to be a good predictor of diversity in the interference selection regime, with more than an order of magnitude variation in $\pi/\pi_0$ for the same value of $\lambda$. However, when the same populations are reorganized according to their variance in fitness (\fig{fig:t2-collapse}B), the pattern essentially flips. The variance in fitness within the population is a strikingly accurate predictor for $\pi/\pi_0$ in the interference selection regime (circles), but it is a poor predictor in the background selection regime (triangles).

The corresponding distortions in the site frequency spectrum are illustrated in \fig{fig:nonparametric-stats-collapse}. For clarity, we only include populations in the interference selection regime, and we focus on two summaries of the full site frequency spectrum.\footnote{The full site frequency spectra are shown in Fig.~S5.} \Fig{fig:nonparametric-stats-collapse}A shows the excess of rare alleles as measured by the reduction in average minor allele frequency. These distortions cannot be explained by \emph{any} constant $N_e$, including the background selection limit in \eq{eq:bgs-limit}. Similarly, \fig{fig:nonparametric-stats-collapse}B shows a measure of the non-monotonic or ``U-shaped'' dependence at high frequencies, using the statistic $\min_i \log \left[ S_n(i) / S_n(n-1) \right]$. In this case, deviations from neutrality reflect topological properties of the genealogy, and they cannot be explained by \emph{any} time-dependent $N_e(t)$ or any exchangeable coalescent model [e.g., \citep{ofallon:etal:2010}] that ignores multiple mergers \citep{sargsyan:wakeley:2008}. \Fig{fig:nonparametric-stats-collapse} shows that even simple models of purifying selection can lead to strong distortions of the frequency spectrum at silent sites, and that these distortions can persist even when individual mutations are only weakly selected against ($Ns \sim 1$).

Yet the most striking feature of these distortions is not simply that they exist, but rather that they are extremely well-predicted by the reduction in pairwise diversity in these populations --- which is itself well-predicted by the variance in fitness. This strong correlation is a nontrivial feature of interference selection, and it disappears for the populations that were classified into the background selection regime (Fig.~S2). \Fig{fig:nonparametric-stats-collapse} also shows that correlations persist when we repeat our simulations with nonzero rates of recombination, suggesting that there is nothing inherently singular about the $R \approx 0$ limit. As long as a sufficient density of selected mutations is maintained, recombination seems to modify only the \emph{degree} of the distortions from neutrality, while the qualitative nature of the distortions remains the same.

Together, \figs{fig:t2-collapse}{fig:nonparametric-stats-collapse} suggest an approximate ``collapse'' or reduction in dimensionality from the original four-parameter model to a single-parameter curve. This carries a number of practical benefits for prediction, which we will exploit to our advantage below. However, this increased predictive capacity places fundamental limits on our ability to resolve individual selection pressures from patterns of silent site variability, even in this highly idealized setting. Our simulations show that two asexual populations with the same variance in fitness will display nearly identical patterns of silent site variability in the interference selection regime, regardless of the fitness effects of the nonsynonymous mutations.

\section{Analysis}

The purpose of this section is twofold. First, we will use a heuristic mathematical argument to show that the patterns observed in the previous section are a natural consequence of our model in the limit of large mutation rates. Then, we will show how this limit can be leveraged to \emph{predict} patterns of genetic diversity, by establishing a mapping between these populations and those in the background selection regime. This mapping allows us to use existing theory to obtain efficient and accurate predictions when interference selection is common.

We focus first on the asexual case where $R = 0$. This leads to a key simplification: different genotypes with the same fitness are completely equivalent, both in terms of their reproductive capacity and their potential for future mutations. The evolutionary dynamics is completely determined by the proportion, $f(X)$, of individuals in each \emph{fitness class} $X$. The frequency of a mutant allele at some particular site can be modeled in a similar way, by partitioning $f(X)$ into the contributions $f_0(X)$ and $f_1(X)$ from the ancestral and derived alleles. These fitness classes obey the Langevin dynamics
\begin{equation}
\begin{aligned}
\label{eq:stochastic-mutation-langevin}
\frac{\partial f_i(X)}{\partial t} & = \left[ X-\overline{X}(t) \right] f_i(X) + U \left[ f_i(X+s) - f_i(X) \right] \\
	& \quad + \sum_{j,X'} \left[ \delta_{ij} \delta_{X X'} - f_i(X) \right] \sqrt{\frac{f_j(X')}{N}} \eta_j(X') \, ,
\end{aligned}
\end{equation}
where $\overline{X}$ is the mean fitness of the population, and $\eta_i(X)$ is a Brownian noise term \citep{good:desai:2013}. \Eq{eq:stochastic-mutation-langevin} tracks only the \emph{fitnesses} of the mutant offspring as they accumulate additional mutations, and it represents a natural extension of the diffusion limit for genomes with a large number of selected sites.

\subsection{The background selection limit}

\noindent \newline The advantage of the description in \eq{eq:stochastic-mutation-langevin} is that it can be analyzed with standard perturbative techniques. For example, while background selection is not always motivated in this fashion, \eq{eq:bgs-limit} arises as a formal limit of the dynamics in \eq{eq:stochastic-mutation-langevin} when $Ns \to \infty$ (SI Appendix). To avoid the trivial behavior $\pi/\pi_0 \to 1$, where selection can be entirely neglected, we must also take $NU \to \infty$ so that the deleterious load $\lambda$ (and therefore $\pi/\pi_0$) remains constant. In this limit, molecular evolution is completely determined by $\lambda$, or equivalently by $N_e/N$, which represents the fraction of mutation-free individuals in the population.

Of course, selection strengths are never infinite in practice. The collapse in \fig{fig:t2-collapse}A --- and the utility of background selection more generally --- is a product of \emph{convergence} to the background selection limit when $N_e s$ is large but finite. In some sense, it is remarkable that \eq{eq:bgs-limit} provides such a good approximation in \fig{fig:t2-collapse}A even when $N_e s$ is as low as $N_e s \sim 10$. However, while \eq{eq:bgs-limit} provides a good approximation for $\pi/\pi_0$, deviations from the background selection limit can be observed in other quantities such as the site frequency spectrum (Fig.~S2). Fortunately, provided $N_e s$ is not too small, we can predict these distortions using a \emph{perturbative approach}, which treats them as (initially) small corrections to the background selection limit. The \emph{structured coalescent} is a widely used method for calculating key aspects of these corrections; we describe it in more detail in the SI Appendix.

The present discussion emphasizes that background selection and the structured coalescent ultimately describe a ``strong selection'' ($N_e s \gg 1$) limit, although the precise meaning of strong is somewhat different from colloquial usage. In particular, this is not just a statement about the magnitude of $Ns$ alone. Selection strengths that are considered strong in a single-site setting ($Ns \sim 30$) can nevertheless have $N_e s \ll 1$ if their collective mutation rate is sufficiently high (e.g., the \emph{D. melanogaster} dot chromosome in \fig{fig:frequency-spectrum-diagram}). Nor is this simply a statement about the magnitude of $U/s$.  Somewhat confusingly, background selection is sometimes regarded as a ``weak selection'' effect, since $N_e$ is significantly reduced only when $s \lesssim U$. We will avoid such terminology here. Instead, it is useful to think of background selection as a ``rare interference'' limit, in contrast to the regime of strong interference analyzed below. 

\begin{figure*}[t]
\centering
\includegraphics[width=0.95\textwidth]{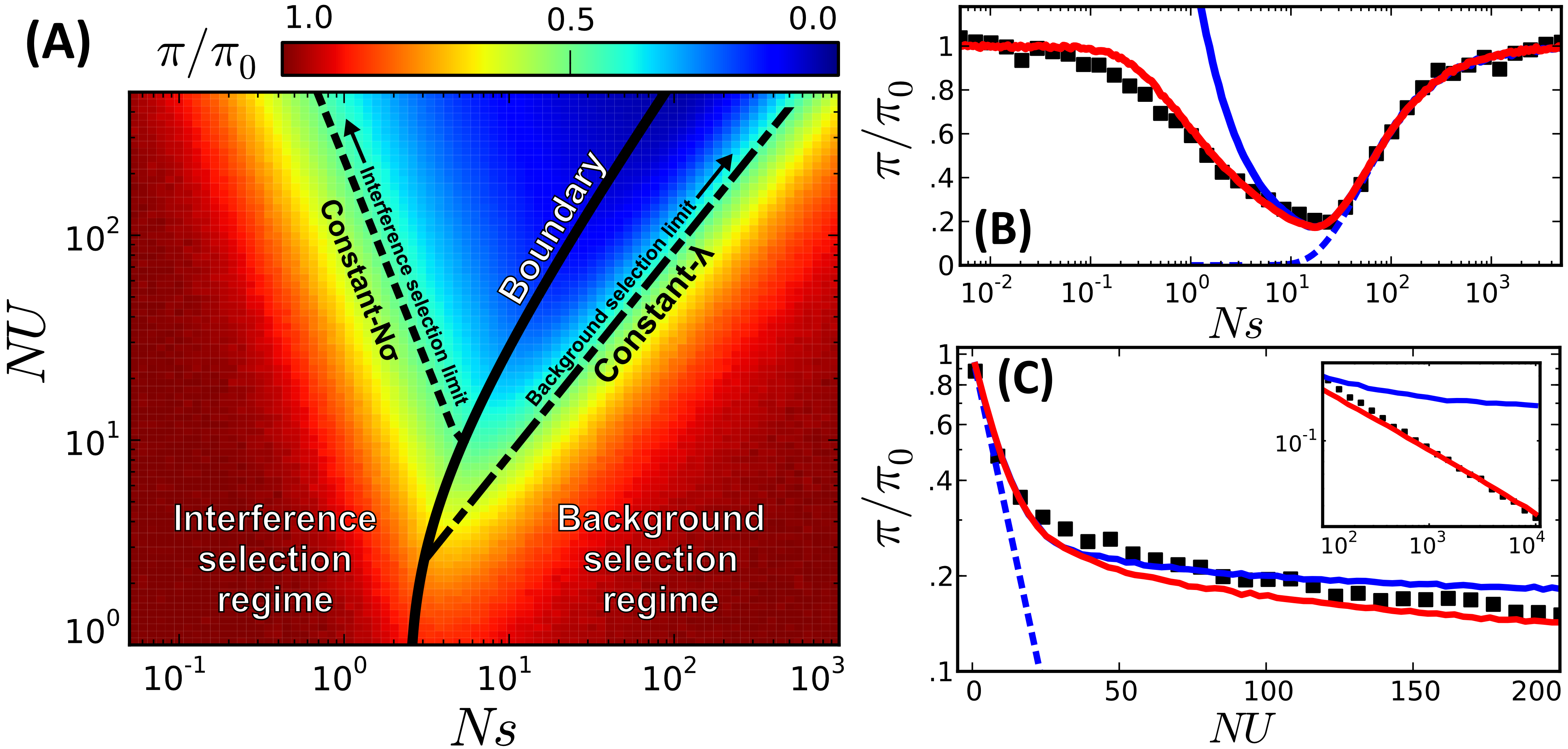}
\vspace{1em}
\caption{(A) Predictions for the reduction in pairwise diversity obtained from our coarse-grained coalescent framework. The solid black line denotes the boundary separating the interference and background selection regimes, while the dashed lines to the left and right denote lines of constant $N\sigma$ and lines of constant $\lambda$, respectively. (B) A ``slice" of this phase plot for constant $NU = 50$. The black squares denote the results of forward-time simulations and our coarse-grained predictions are shown in solid red. For comparison, the original structured coalescent is shown in solid blue, while the dashed line gives the prediction from the background selection limit in \eq{eq:bgs-limit}. (C) A similar ``slice'' of this phase plot for constant $\Ns = 10$. As $NU \to \infty$, we approach the asymptotic limit $\pi/\pi_0 \sim (NU)^{-1/3}$ from \citeref{neher:hallatschek:2013}. \label{fig:t2}}
\end{figure*}

\subsection{The interference selection limit}

\noindent \newline Since the collapse in \fig{fig:t2-collapse}A is driven by the background selection limit, we might conclude that \fig{fig:t2-collapse}B is the result of a second, \emph{interference selection limit} when $Ns \to 0$. Such a limit would be as fundamental for the interference regime as \eq{eq:bgs-limit} was for the background selection regime above. Of course, if $Ns$ vanishes on its own we simply recover the neutral limit, $\pi/\pi_0 \to 1$. To maintain nontrivial behavior, \fig{fig:t2-collapse}B suggests that we take $NU \to \infty$ as well, so that the variance in fitness (and therefore $\pi/\pi_0$) remains constant. In this way, interference selection resembles a \emph{linked} version of the infinitesimal trait models from quantitative genetics, where phenotypic (fitness) variation arises from a large number of small-effect (weakly-selected) alleles.

While the evidence from \fig{fig:t2-collapse}B is suggestive, we can establish the interference selection limit more rigorously using \eq{eq:stochastic-mutation-langevin}, where it corresponds to the limit that $Ns \to 0$ and $NU \to \infty$ with the product $N^3 U s^2$ held constant (SI Appendix). Extension to a distribution of fitness effects is straightforward, provided that we replace $s^2$ with $\langle s^2 \rangle = \int s^2 \rho(s) \, ds$. In this limit, the distribution of fitnesses within the population and the patterns of molecular evolution depend only on the product $N^3 U\langle s^2 \rangle$ and not any other properties of $\rho(s)$. In particular, the effects of beneficial and deleterious mutations are symmetric. Selected mutations are negligible on their own, and are virtually indistinguishable from neutral mutations, but the population as a whole is far from neutral. Rather, infinitesimal mutations arise so frequently that the population maintains substantial variation in fitness, and this leads to correspondingly large distortions at the sequence level. The distribution of fitnesses within these populations is well-characterized by ``traveling wave'' models of fitness evolution \citep{tsimring:etal:1996, cohen:kessler:levine:2005, hallatschek:2011}, which provide explicit formulae for the variance in fitness as a function of the control parameter $N^3 U \langle s^2 \rangle$ (SI Appendix). These formulae show that $N\sigma$ increases monotonically with $N^3 U \langle s^2 \rangle$, so either quantity can be used to index populations in the interference selection limit. We will use $N\sigma$ for the remainder of the paper in order to maintain consistency with \fig{fig:t2-collapse}. Due to interference, this variance is typically much smaller than the deterministic prediction, $\sigma_\mathrm{det}^2 = Us$, attained under background selection. 

Unfortunately, patterns of molecular evolution are less well-characterized in this limit, which makes it difficult to \emph{predict} the correlations in \figs{fig:t2-collapse}{fig:nonparametric-stats-collapse}. Like the background selection limit, we would then hope that this genealogical description remains valid for finite $NU$ and $Ns$, so that we could use the interference selection limit to obtain predictions for biologically relevant parameters. 
However, a complete description has been obtained only in the special case where $N\sigma~\to~0$ or $N\sigma~\to~\infty$. The former corresponds to a neutral population, with small corrections calculated in \citeref{good:desai:2013}. In the latter case, the genealogy of the population approaches that of the Bolthausen-Sznitmann coalescent \citep{bolthausen:sznitman:1998}, and distortions in the site-frequency spectrum are independent of \emph{all} underlying parameters \citep{neher:hallatschek:2013}. However, \fig{fig:nonparametric-stats-collapse} shows that many biologically relevant parameters fall somewhat far from the $N\sigma \to \infty$ limit, so we require an alternate method to predict genetic diversity for the moderate values of $N\sigma$ that are likely to arise in practice.

\subsection{Predicting genetic diversity by coarse-graining fitness}

\noindent \newline In the absence of a ``fine-grained'' solution, we employ an alternate strategy inspired by the simulations in \figs{fig:t2-collapse}{fig:nonparametric-stats-collapse}. Convergence to the interference selection limit is \emph{extremely} rapid in these figures --- so rapid that we can effectively neglect any corrections to this limit all the way up to the boundary of the background selection regime. In other words, the structured coalescent and the interference selection limit are \emph{both} approximately valid along this boundary. Rather than using the ``fine-grained'' limit to approximate populations with a given $N\sigma$, this rapid convergence suggests that we could also use a population on the \emph{boundary} of the background selection regime with the same $N\sigma$. Intuitively, this resembles a ``coarse-graining'' of fitness distribution, since it approximates several weakly selected mutations in the original population with a smaller number of strongly selected mutations in the background selection regime. On a formal level, this is nothing but a \emph{patching method} \citep{bender:orszag:1978} that connects the asymptotic behavior in the interference selection ($Ns \to 0$) and background selection ($Ns \to \infty$) limits. 

This intuition suggests a simple algorithm for predicting genetic diversity in the interference selection regime: (i)~calculate $N\sigma$ as a function of $Ns$ and $NU$ as described in SI Appendix, (ii)~find a corresponding point on the boundary of the background selection regime with the same $N\sigma$, and (iii)~evaluate the structured coalescent using these ``coarse-grained'' parameters. Step (ii) requires a precise definition of the boundary between the interference and background selection regimes, which we have not yet specified. Like many patching methods, this boundary is somewhat arbitrary, since the transition between the $Ns \to 0$ and $Ns \to \infty$ limits is not infinitely sharp. Our definition here is based on a specific feature of the structured coalescent: for each $N\sigma$, this method starts to break down near the point of \emph{maximum} reduction in pairwise diversity (SI Appendix), which is close to the crossover point where Muller's ratchet starts to click more frequently \citep{gordo:etal:2002}. Together, these maxima define a ``critical line'' in the $(Ns,NU)$ plane (\fig{fig:t2}A), which serves as the boundary between the interference and background selection regimes. Populations above or to the left of this line are classified into the interference selection regime, and the genetic diversity in these populations is calculated using the algorithm above. The remaining populations belong to the background selection regime, where the structured coalescent is already valid.

This procedure rapidly generates predictions across the full range of parameters, as illustrated in \fig{fig:t2} for the pairwise diversity, $\pi/\pi_0$. We see that the coarse-grained predictions accurately recover the transition to the neutral limit when $Ns \to 0$ (\fig{fig:t2}B), and they also reproduce the power-law decay in heterozygosity when $NU \to \infty$ (\fig{fig:t2}C). Similar predictions in \fig{fig:nonparametric-stats-collapse} reproduce the distortions in the site-frequency spectrum as well.

\subsection{Emergence of linkage blocks in recombining genomes}

\noindent \newline So far, our analysis has focused on nonrecombining genomes, but our simulations in \fig{fig:nonparametric-stats-collapse} show that similar behavior arises when $R > 0$ as well. Yet a formal analysis is more difficult in this case. Recombination requires explicit haplotype information and cannot be recast in terms of the evolution of fitness alone. Thus, while the structured coalescent can be extended to recombining genomes \citep{zeng:charlesworth:2011}, there is no simple analogue of \eq{eq:stochastic-mutation-langevin} that we can use to \emph{formally} extend the interference selection limit.

Nevertheless, we can gain considerable insight with a simple heuristic argument, which leverages our previous analysis in nonrecombining genomes. Neighboring regions of a linear chromosome recombine much less than the genome as a whole. Sites separated by a map length $\Delta R \ll 1/T_\mathrm{MRCA}$ will typically not recombine at all in the history of the sample, so the ancestral process should predominantly resemble an asexual population on these length scales. In the opposite extreme, sites with $\Delta R \gg 1/T_\mathrm{MRCA}$ will recombine many times in the history of the sample, and will effectively act as if they were unlinked. To the extent that this transition is sharp, the evolution of a recombining genome can be viewed as a collection of independent, freely recombining \emph{linkage blocks}, each of which evolves asexually. Intuitively, this resembles the ``sliding window'' techniques used to speed up simulation and data analysis in large genomes.

If each block comprises a fraction $L_b/L$ of the genome, then the distribution of fitness and the patterns of molecular evolution within each block are by definition the same as an asexual population with an effective mutation rate
\begin{align}
\label{eq:ueff}
U_{\mathrm{eff}} = \left( \frac{L_b}{L} \right) U \, .
\end{align}
Technically, the unlinked blocks also contribute to a reduction in the effective population size \citep{neher:shraiman:qtl:2011}, but we neglect these effects here (see SI Appendix). The block size itself must satisfy the condition that there are few recombination events within a block in a typical coalescence time, or
\begin{align}
\label{eq:linkage-block}
R \left( \frac{L_b}{L} \right) \cdot T_2 \sim 1 \, .
\end{align}
Here, $T_2 = N \pi / \pi_0$ is the pairwise coalescence time for the linkage block, which is itself a function of $L_b/L$ and can be calculated from \eq{eq:ueff} and the asexual methods above. Together, \eqs{eq:ueff}{eq:linkage-block} uniquely determine the block size in a given population. Using our coarse-grained predictions for $\pi/\pi_0$, we can solve for $L_b/L$ and obtain explicit predictions for the molecular evolution in recombining genomes. In practice, we use a generalized version of \eq{eq:linkage-block} in SI Appendix that accounts for constant factors and the saturation of the block size when $T_2 R \lesssim 1$. 

\graphicspath{{./reorganized_figures/}}
\begin{figure}[t]
\centering
\includegraphics[width=0.95\columnwidth]{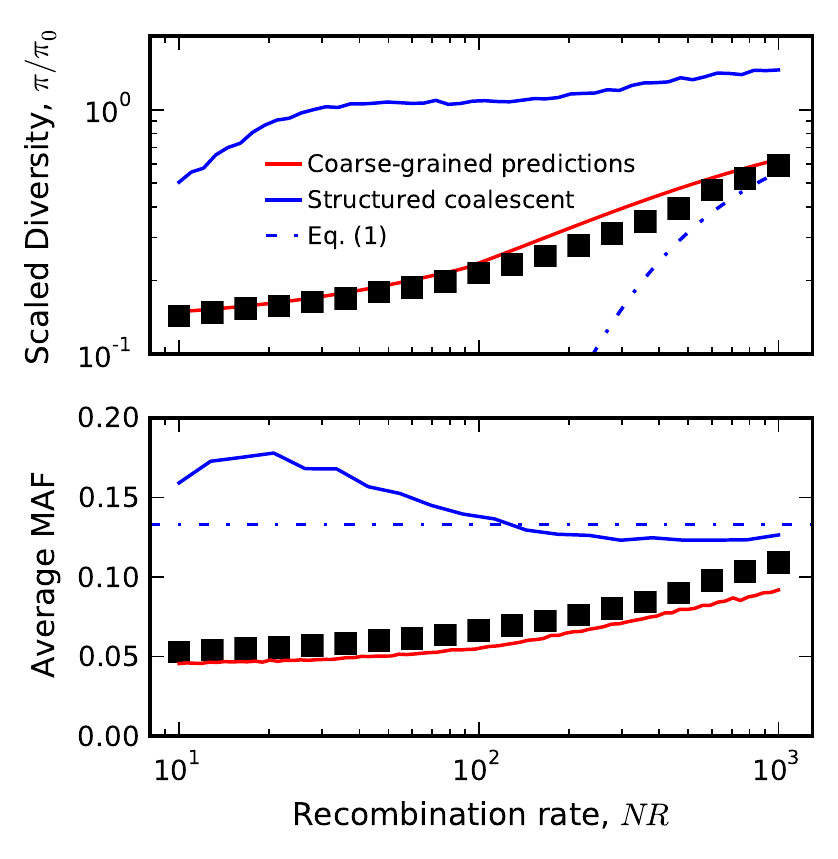}
\vspace{1em}
\caption{Correlation between diversity and recombination rate in the presence of interference. Black squares denote the results of forward time simulations for $Ns=10$ and $NU=300$, while our coarse-grained predictions are shown in solid red. For comparison, we have also included predictions from the background selection limit in \eq{eq:bgs-limit} as well as the recombinant structured coalescent predictions from \citep{zeng:charlesworth:2011}.\label{fig:diversity-correlation}}
\end{figure}

\Fig{fig:diversity-correlation} shows an example of these predictions for the correlation between diversity and recombination rate, which is often used to infer selection in natural populations. Our simple approximation is surprisingly accurate for these parameters, especially compared to the background selection limit in \eq{eq:bgs-limit} or the recombining structured coalescent in \citeref{zeng:charlesworth:2011}. This accuracy is especially surprising given that the predictions are obtained from an \emph{asexual} population with a coarse-grained selection strength and mutation rate. Evidently, interference on a linear chromosome more closely resembles an asexual genome (with an appropriately defined length) rather than the freely recombining, single-site models that are more commonly employed. A more thorough investigation of the linkage block concept and its implications for other aspects of sequence diversity (e.g., linkage disequilibria, variation in recombination rate, etc.) remain an important avenue for future work.

\section{Discussion}

Interfering mutations display complex dynamics that have been difficult to model with traditional methods. Here, we have shown that simple behavior emerges in the limit of widespread interference. When fitness variation is composed of many individual mutations, the magnitudes and signs of their fitness effects are relatively unimportant. Instead, molecular evolution is controlled by the variance in fitness within the population over some effectively asexual segment of the genome. This implies a corresponding \emph{symmetry}, in which many weakly selected mutations combine to mimic the effects of a few strongly deleterious mutations with the same variance in fitness. We have exploited this symmetry in our ``coarse-grained'' coalescent framework, which generates efficient predictions across a much broader range of selection pressures than was previously possible.

Of course, the degree of interference in any particular organism is ultimately an empirical question --- one that hinges on the relative strengths of mutation, selection, and recombination. Although interference is often observed in microbes \citep{miralles:etal:1999, kao:sherlock:2008, strelkowa:laessig:2012},  its prevelance in higher sexual organisms is still controversial because it is difficult to estimate these parameters in the wild. Mutation and recombination rates can be measured directly (at least in principle), but population sizes and selection strengths can only be \emph{inferred} from a population genetic model, and these have historically struggled to include the effects of selection on linked sites. Many estimates of ``$N_e s$'' ignore linkage by fiat (e.g. \citep{loewe:charlesworth:2006}) under the assumption that sites evolve independently. But these estimates become unreliable precisely when small- and intermediate-effect mutations are most common, and the reasons for this are apparent from \fig{fig:nonparametric-stats-collapse}. All of the distortions in \fig{fig:nonparametric-stats-collapse}A would be mistakenly ascribed to demography (or in the case of \fig{fig:nonparametric-stats-collapse}B, population substructure), thereby biasing the estimates of selection at nonsynonymous sites. At best, these estimates of ``$N_e s$'' represent measurements of $\T2 s$, which carries little information about the true strength of selection ($Ns$) or even the potential severity of interference. For example, all of the populations in \fig{fig:diversity-correlation} have $Ns = 10$ and $\T2 s > 1$, even though they show a strong distortion in minor allele frequency that cannot be explained by \eq{eq:bgs-limit}. In other words, we cannot conclude that interference is negligible just because ``$N_e s$'' is larger than one. 

\begin{figure}[t]
\centering
\includegraphics[width=0.95\columnwidth]{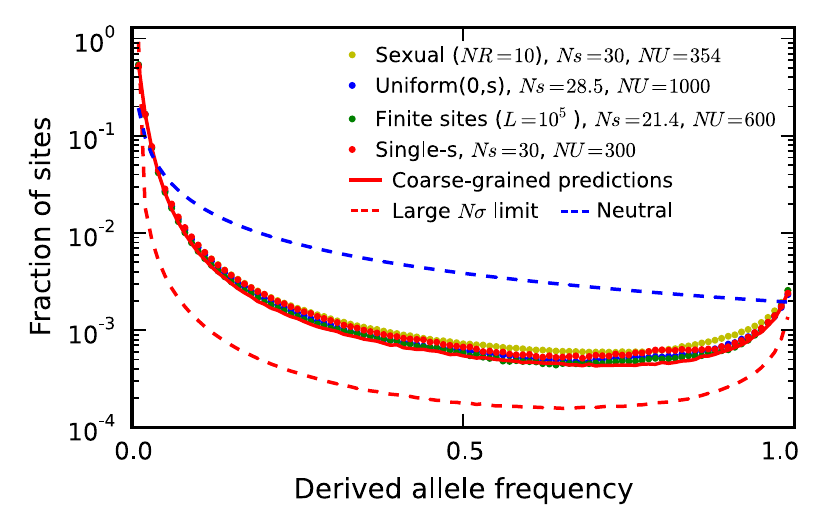}
\caption{The site frequency spectrum from \fig{fig:frequency-spectrum-diagram} (red dots) and forward-time simulations of three equivalent populations predicted from our coarse-grained theory: a recombining population (yellow), a finite chromosome with $L=10^5$ sites that allows for beneficial as well as deleterious mutations (green), and a population with a uniform distribution of deleterious fitness effects (blue). Our coarse-grained predictions are shown in solid red. For comparison, the dashed blue lines show the neutral expectation, while the dashed red lines show the large $N\sigma$ limit \citep{neher:hallatschek:2013}. \label{fig:frequency-collapse}}
\end{figure}

More sophisticated analyses avoid these issues with simulations of the underlying genomic model \citep{kaiser:charlesworth:2008, seger:etal:2010, lohmueller:etal:2011}. In principle, this approach can provide robust estimates of the underlying parameter combinations that best describe the data. But in practice, simulation-based methods suffer from two major shortcomings which are highlighted by the symmetry above. We have seen that strongly-interfering populations with the same variance in fitness possess nearly identical patterns of genetic diversity. This suggests a degree of ``sloppiness'' \citep{gutenkunst:etal:2007} in the underlying model, which can lead to large intrinsic uncertainties in the parameter estimates and a strong sensitivity to measurement noise. A more fundamental problem is identifying the nearly equivalent populations in the first place. Even in our simplified model, large genomes are computationally expensive to simulate, and this obviously limits both the number of dependent variables and the various parameter combinations that can be explored in a single study. We have shown that sets of equivalent populations lie along a single line (namely, the line of constant $N\sigma$) in the larger parameter space, which can easily be missed in a small survey unless the parameters are chosen with this degeneracy in mind. In this way, our theoretical predictions can aid existing simulation methods by identifying equivalent sets of parameters that also describe the data. 

As an example, we consider the \emph{D. melanogaster} dot chromosome that inspired the parameter combination in \fig{fig:frequency-spectrum-diagram}. Earlier, we showed that the silent site diversity on this chromosome ($\pi/\pi_0 \sim 7\%$) is consistent with the parameters $Ns \approx 30$, $NU \approx 300$, and $NR \approx 0$, which fall in the middle of the interference selection regime (\citep{kaiser:charlesworth:2008}, SI Appendix). Our calculations allow us to predict other parameter combinations with the same patterns of diversity, and we plot the simulated frequency spectrum for three of these alternatives in \fig{fig:frequency-collapse}. We see that even with highly resolved frequency spectra (unavailable in the original dataset), there is little power to distinguish between these predicted alternatives despite rather large differences in the underlying parameters. 

However, this ``resolution limit'' suggests that individual fitness effects are not the most interesting quantity to measure when interference is common. Individual fitness effects may play a central role in single-site models, but we have shown that global properties like the variance in fitness and the corresponding linkage scale are more relevant for predicting evolution in interfering populations. Estimating these quantities directly may therefore be preferable in practice. Our coarse-grained predictions provide a promising new framework for inferring these quantities based on allele frequency data \citep{sawyer:hartl:1992} or genealogical reconstruction \cite{huelsenbeck:ronquist:2001}. A concrete implementation presents a number of additional challenges, mostly to ensure a proper exploration of the high-dimensional parameter space, but this remains an important avenue for future work.

Finally, our findings suggest a \emph{qualitative} shift in the interpretations gleaned from previous empirical studies. We have provided further evidence that even weak purifying selection, when aggregated over a sufficiently large number of sites, can generate strong deviations from neutrality. Moreover, these signals can resemble more ``biologically interesting'' scenarios like recurrent sweeps or large-scale demographic changes. Here we refer not only to the well-known reduction in diversity and skew towards rare alleles, but also to the topological imbalance in the genealogy (or the ``U-shaped'' frequency spectrum), and the strong correlations in these quantities with the rate of recombination. Since weakly deleterious mutations are already expected to be common \citep{eyre-walker:keightley:2007}, they may constitute a more parsimonious explanation for observed patterns of diversity unless they can be rejected by a careful, quantitative comparison of the type advocated above. At the very least, these signals should not be interpreted as \emph{prima facie} evidence for anything more complicated than weak but widespread purifying selection.

\begin{acknowledgments}
We thank Asher Cutter, John Wakeley, Lauren Nicolaisen, Elizabeth Jerison, Daniel Balick, and Katya Kosheleva for comments and useful discussions. This work was supported in part by a National Science Foundation Graduate Research Fellowship (B.H.G.), the James S. Mcdonnell Foundation and the Harvard Milton Fund (M.M.D), and the European Research Council Grant no. 306312 (A.M.W.). Simulations in this paper were performed on the Odyssey cluster supported by the Research Computing Group at Harvard University. 
\end{acknowledgments}

\end{document}